\documentstyle[prd,aps,preprint]{revtex}

\def\sqr#1#2{{\vcenter{\hrule height.#2pt
   \hbox{\vrule width.#2pt height#1pt \kern#1pt
    \vrule width.#2pt} 
    \hrule height.#2pt}}}
\def\dalamb {\mathchoice{\sqr{6.9}4}{\sqr{6.9}4}{\sqr{5.5}3}{\sqr{4.6}3}}
\def\meio { \frac{1}{2} }
\def\Lie { {\cal L}_{\xi} }

\def\tQ { \tilde{\cal Q} }

\def\tP { \tilde{\cal P} }
\def\tq { \tilde{q} }
\def\tg { \tilde{g} }

\def\tx { \tilde {x} }

\def\taumn { \tau_{\mu \nu} }
\def\lphir { \langle \phi^2 \rangle }
\newcommand{\be}{\begin{equation}}
\newcommand{\ee}{\end{equation}}
\newcommand{\beq}{\begin{eqnarray}}
\newcommand{\eeq}{\end{eqnarray}}

\begin{document}


\title{The Energy-Momentum Tensor for Cosmological Perturbations}
\author{L. Raul W. Abramo$^1 \,$\cite{maila}, Robert H. 
Brandenberger$^1 \, $\cite{mailb} \\
and V. F. Mukhanov$^2 \, $\cite{mailc}}
\address{$^1$ Physics Department, Brown University,\\
Providence, R.I. 02912, USA;\\
$^2$ Institut f\"ur Theoretische Physik, ETH Z\"urich, CH-8093 \\
Z\"{u}rich, Switzerland}
\maketitle

\begin{abstract}
\noindent
We study the effective energy-momentum tensor (EMT) for cosmological
perturbations and formulate the gravitational back-reaction problem in a
gauge invariant manner. We analyze the explicit expressions for the EMT in
the cases of scalar metric fluctuations and of gravitational waves and
derive the resulting equations of state. 

The formalism is applied to
investigate the back-reaction effects in chaotic inflation.
We find that for long wavelength scalar and tensor perturbations, the
effective energy density is negative and thus counteracts any pre-existing
cosmological constant. For scalar perturbations during an epoch of
inflation, the equation of state is de Sitter-like.

\end{abstract}

\vskip 0.8cm \noindent {PACS numbers: 98.80Cq\\BROWN-HET-1046\\April 1997}

\newpage

\section{Introduction}

It is well known$^{\cite{Landau,MTW,Weinb}}$ that gravitational waves
propagating in some background space-time have an effect on the dynamics of
this background. A convenient way to describe the back reaction of the
fluctuations on the background is in terms of an effective energy-momentum
tensor (EMT). In the short wave limit, when the typical wavelength of
gravitational waves is small compared with the curvature of the background
space-time, they act as a radiative fluid with an equation of state 
$p={\frac 13}\,\rho $ (where $p$ and $\rho $ denote pressure and energy
density, respectively). The presence of a large amount of gravity waves in
the early Universe can lead to important consequences for cosmology. For
example, it can speed up nucleosynthesis and lead to a higher fraction of
helium, resulting in constraints on models producing a too large amplitude
of gravitational waves in the early Universe (see e.g. Ref. \cite{nucleo}
and references quoted therein).

In most models of the early Universe, scalar-type metric perturbations are
more important than gravity waves. On length scales smaller than the Hubble
radius, the amplitude of scalar fluctuations increases in time, and in most
models, scalar perturbations are responsible for seeding structure in the
Universe. In this paper, we study the back reaction problem for both scalar
and tensor perturbations (cosmological perturbations and gravitational
waves, respectively). We derive the effective EMT which describes the
back-reaction and apply the result to calculate EMT for both long and short
wavelength fluctuations in particular models for the evolution of the Universe.

One of the main puzzles to be solved is the problem of gauge invariance of
the effective EMT. As is well known (see e.g. Ref. \cite{MFB92} for a
comprehensive review), cosmological perturbations transform non-trivially
under coordinate transformations (gauge transformations). However, the
answer to the question ``how important are perturbations for the evolution
of a background'' must be independent of the choice of gauge, and hence the
back-reaction problem must be formulated in a gauge invariant way.

In a recent Letter$^{\cite{MAB96}}$, we demonstrated how the back reaction
problem can be set up in a gauge invariant manner. We applied the result to
estimate the magnitude of back reaction effects in the chaotic inflationary
Universe scenario. In this paper, we study in more detail the effective EMT
of cosmological perturbations. In particular, we derive the equation of
state satisfied by this EMT. As we show, the back reaction effects of
gravity waves and of scalar fluctuations decouple. In the short wavelength
limit, we recover the result $p={\frac 13}\,\rho $ for gravity waves. In
the long wavelength limit, scalar fluctuations about a de Sitter background
have an equation of state $p\approx -\rho $ with $\rho < 0$.

The study of back-reaction effects for gravitational waves goes back a long
way. Following pioneering work of Brill and Hartle$^{\cite{Geon}}$, 
Isaacson$^{\cite{Isaacson}}$ defined an effective EMT for gravitational
waves which
was shown to be gauge-invariant for high-frequency waves after averaging
over both space and time. This prescription only makes sense, however, when
considering fluctuations on scales much smaller than those characterizing
the background. In applications to physics of the very early Universe
the fluctuations of interest have wavelengths larger than the
Hubble radius and a frequency smaller than the expansion rate. Hence,
Isaacson's procedure for defining an effective EMT is inapplicable.

Back reaction effects for density inhomogeneities have been considered only
recently, and even then without addressing questions of gauge-dependence.
The focus of the early work of Futamase et al$^{\cite{Futamase}}$ and of
Seljak and Hui$^{\cite{Seljak}}$ was on effects of inhomogeneities on local
observables such as the expansion rate of the Universe. For a recent study
of this issue in the context of Newtonian cosmology, the reader is referred
to the work of Buchert and Ehlers$^{\cite{B-E}}$. The focus of our work, on
the other hand, is to formulate the back-reaction problem in General
Relativity in a gauge-invariant manner.

The outline of this paper is as follows. In the following section we
formulate some useful properties of diffeomorphism transformations. The back
reaction problem is set up in Section 3 and then in the next section we
recast the back reaction problem in terms of gauge invariant variables. In
Section 5 we first demonstrate that the contributions of scalar and tensor
fluctuations to the effective EMT do not interfere in the leading
approximation. We then study in detail the effective EMT for scalar
perturbations, focusing on the equations of state obtained in the long and
short wavelength limits. In Section 6 we derive the effective EMT for
gravitational waves. As an application, we consider the back reaction of
cosmological perturbations in the chaotic inflationary Universe scenario. We
summarize our results in Section 8.

\vskip .20in

\section{Gauge Transformations}

The gauge group of general relativity is the group of diffeomorphisms. A
diffeomorphism corresponds to a differentiable coordinate transformation.
The coordinate transformation on the manifold ${\cal {M}}$ can be
considered as generated by a smooth vector field $\xi^\alpha $. 
Let us take some coordinate system $x^\alpha$ on ${\cal {M}}$ in which for
instance some arbitrary point $P$ of that manifold has 
coordinates $x_P^\alpha $. The solution of the differential equation

\begin{equation}
{\frac{{d\chi ^\alpha (P;\lambda )}}{{d\lambda }}}=\xi ^\alpha \left[ \chi
(P;\lambda )\right] \,,  \label{flow}
\end{equation}
with initial conditions

\begin{equation}
\chi ^\alpha (P;\lambda =0)=x_P^\alpha
\end{equation}
defines the parametrized integral curve $x^\alpha (\lambda )=\chi ^\alpha
(P;\lambda )$ with the tangent vector $\xi ^\alpha (x_P)$ at $P$. Therefore,
given the vector field $\xi ^\alpha $ on ${\cal {M}}$ we can define an
associated coordinate transformation on ${\cal {M}}$ as, for instance, 
$x_P^\alpha \stackrel{\xi}\rightarrow \tilde x_P^\alpha = 
\chi ^\alpha (P;\lambda =1)$ for any given $P$.
Assuming that $\xi ^\alpha $ is small one can use the 
perturbative expansion for
the solution of equation (\ref{flow}) to obtain

\begin{equation}
\tilde x_P^\alpha = 
\chi^\alpha (P;\lambda =1) = x_P^\alpha + \xi^\alpha (x_P) + {\frac 12} 
\xi_{,\beta }^\alpha \xi^\beta + {\cal {O}}(\xi^3) \,  ,\label{ct}
\end{equation}
which we can write in short-hand notation as

\begin{equation}
\chi _P^\alpha (P;\lambda =1)=(e^{\xi ^\beta \frac \partial {\partial
x^\beta }}x^\alpha )_P  \label{ct2}
\end{equation}
Thus the general coordinate transformation 
$x\stackrel{\xi }{\rightarrow } \tilde x$ 
on ${\cal {M}}$ generated by the vector field $\xi ^\alpha (x)$
can be written as 
\begin{equation}
x^\alpha \rightarrow \tilde x^\alpha =e^{\xi ^\beta \frac \partial {\partial
x^\beta }}x^\alpha   \label{coor}
\end{equation}
Conversely, given any two coordinate systems $x^\alpha $ and $\tilde x
^\alpha $ on ${\cal {M}}$ which are not too distant, we can find the vector
field $\xi ^\alpha (x)$ which generates the coordinate transformations $
x\rightarrow \tilde x$ in the sense (\ref{coor}). Of course, in such a way
we can not cover all possible coordinate
transformations$^{\cite{Knight}}$. However, the class
of transformations described above is wide enough for our purposes.

The variables which describe physics on the manifold ${\cal {M}}$ are tensor
fields $Q$. Under coordinate transformations $x\rightarrow \tilde x$ the
value of $Q$ at the {\it given point }$P$ {\it of the manifold }transforms
according to the well known law
\begin{equation}
Q(x_P)\rightarrow {\tQ}(\tx_P)=
{\partial \tilde x \overwithdelims() \partial x}_P \cdots
{\partial x \overwithdelims() \partial \tilde x}_P Q(x_P)  \label{tran}
\end{equation}
Note that both sides in this expression refer to the same point of manifold
which has different coordinate values in different coordinate frames,
that is ${\tx}_P\neq x_P.$ The question about the transformation
law for the tensor field $Q$ can be formulated in a different way. Namely,
given
two different points ${\cal P}$ and ${\tP}$, which have the same
coordinate values in different coordinate frames, that is 
${\tx}_{\tP}=x_{\cal P}$, we could ask how to express the components of the
tensor in the coordinate frame $\tilde x$ at the point $\tP$
(denoted by ${\tQ}_{\tP}$) in terms of $Q$
and its derivatives given in the frame $x$ at the point ${\cal P}$. 
The answer to
this question is found with the help of Lie derivatives with respect to
the vector field $\xi $ which generates the appropriate coordinate
transformation $x\rightarrow \tilde x$ according to (\ref{coor}). Its
infinitesimal form is given in some books on General
Relativity (see, e.g. Ref. \cite{Stewart}):

\begin{equation}
{\tQ}(\tx_{\tP}=x_0) \, = \, [ Q - \Lie Q + O(\xi ^2)] (x_{\cal P} = x_0) 
\,\, ,
\label{Lie1}
\end{equation}
where ${\cal L}_\xi $ denotes the Lie derivative with respect to the vector
field $\xi $. Transformation (\ref{Lie1}) is the infinitesimal form of
the gauge transformations of the diffeomorphism group. The finite form of
it is
obtained by exponentiating (\ref{Lie1}):  
\begin{eqnarray}
Q(x) & \rightarrow & \tQ (x)= (e^ {-\Lie}Q)(x)  \nonumber \\
&=&Q(x)-(\Lie Q)(x)+{\frac 12}(\Lie \Lie Q)(x)+{\cal O}(\xi ^3)  \label{Lie2}
\end{eqnarray}
Equations (\ref{Lie1}) and (\ref{Lie2}) are tensor equations, where for
notational convenience, tensor indices have been omitted. Note that,
despite
of the fact that the transformation law (\ref{Lie2}) is the consequence of
transformation law (\ref{tran}), they are different in the following
respect: transformation law (\ref{tran}) is well defined for any tensor
given
only at the point $P$, while (\ref{Lie2}) is defined only for tensor
{\it fields.}

In the derivations which follow below we make substantial use of some
properties
of the Lie derivatives and elements of functional calculus. Therefore
for the convenience of the reader we would like to recall some basic
useful facts from functional analysis and from the theory of Lie derivatives.
Readers not interested in these formal considerations may skip to the
next section.

Let us consider a tensor field $G(x)$ (e.g., Riemann tensor or Einstein
tensor) which is formed from the metric tensor $g(x)$ and its
derivatives 
$\partial g/\partial x,...,$ that is 

\begin{equation}
G(x) \equiv G[\partial /\partial x,g(x)] \,\, .  \label{def}
\end{equation}
Applying the operator $\exp (-\Lie )$ to $G(x)$, we obtain the value of this
tensor, denoted ${\tilde G}(x)$ at the point $\tP$, 
whose coordinates in the new coordinate frame are
$\tx (x) = x$. On the other hand, ${ \tilde G}(x)$ in the
frame 
$\tx$ can also be calculated from the metric tensor 
${\tg}(x)=(\exp (-\Lie) g)(x)$ and its derivatives 
$\partial {\tg}(x) / \partial x, ... $, according to the
prescription (\ref{def}). The
results should coincide and therefore one can conclude (see Ref.
\cite{Thesis} for an explicit proof of this nontrivial fact)
that

\begin{equation}
(e^{-\Lie }G)(x)=G\left[ \frac \partial {\partial x},(e^{-\Lie}g)(x)\right]
\label{Lie3}
\end{equation}
that is the Lie derivative can be taken though the derivatives $\partial
/\partial x$ without ``changing'' them in expressions where these
derivatives are used to build the tensors (e.g., Riemann tensor) out of the
other tensors (e.g., metric tensor).
 
We will be interpreting functions 
$G(x)=G[\partial /\partial x,g(x)]$ 
defined on the manifold ${\cal {M}}$ as
the parametrized set of functionals $G_x$ defined on the space of
functions $g(x^{\prime })$ according to the formula 

\begin{equation}
G_x \equiv G[\partial /\partial x,g(x)]=\int G[\partial /\partial
x^{\prime},g(x^{\prime })]\delta (x-x^{\prime })dx^{\prime }  \,\,
,\label{f}
\end{equation}
where $\delta (x-x^{\prime })$ is the Dirac delta function. Then the
functional derivative $\delta G_x/\delta g(x^{\prime })$ can be defined in
the
standard way 
\begin{equation}
\delta G_x=\int (\delta G_x/\delta g(x^{\prime }))\delta g(x^{\prime
})dx^{\prime } ,  \label{f1}
\end{equation}
where $\delta G_x$ is the change of the functional $G_x$ under
an infinitesimal
variation of $g(x^{\prime })$: $g(x^{\prime })\rightarrow g(x^{\prime
})+\delta g(x^{\prime }).$

If, for instance, $G_x=g(x),$ then 

\begin{equation}
\delta G_x=\delta g(x)=\int \delta (x-x^{\prime })\delta g(x^{\prime
})dx^{\prime } \,\, , \label{f3}
\end{equation}
and comparing this formula with (\ref{f1}) we deduce that 

\begin{equation}
\frac{\delta G_x}{\delta g(x^{\prime })}=\delta (x-x^{\prime }) \,\, . 
\label{f2}
\end{equation}
As another example, consider $G_x=\partial ^2g(x)/\partial x^2$. Using
the definitions (\ref{f}) and (\ref{f1}) one gets 

\begin{equation}
\frac{\delta G_x}{\delta g(x^{\prime })}=\frac{\partial ^2}{\partial
x^{\prime 2}}\delta (x-x^{\prime }) \,\, .  
\label{f4}
\end{equation}

In the following, the functional derivative $F_{xx^{\prime }}=\delta
G_x/\delta g(x^{\prime })$ will be treated as an operator which acts on
the
function $f(x^{\prime })$ according to the rule 
\begin{equation}
F_{xx^{\prime }}*f(x^{\prime })=\int F_{xx^{\prime }}f(x^{\prime
})dx^{\prime }  \label{f5}
\end{equation}
We will also use DeWitt's condensed notation \cite{DeWitt} and assume that
continuous variables $(t,x^i)$ are included in the indexes, e.g., 
$A^\alpha (x^{i},t)=A^{(\alpha ,x^{i },t)}=A^{a},$
where $a$ is used as the collective variable to denote $(\alpha
,x^{i},t).$ In addition we adopt as a natural extension of the Einstein
summation rule that ``summation'' over repeated indexes
also includes integration over appropriate continuous variables, e.g., 

\begin{equation}
A^a B_a=A^{(\alpha ,x,t)}B_{(\alpha ,x,t)}=\sum_\alpha \int A^\alpha
(x,t)B_\alpha (x,t)dxdt  \label{f6}
\end{equation}

We shall write functional derivatives using the following
short hand notation: 
\begin{equation}
\frac{\delta G_{x}}{\delta g(x')} \equiv \frac{\delta G}
{\delta g^{a'}} \equiv G_{,a'}
\label{f7}
\end{equation}
For instance, the useful formula

\begin{equation}
\Lie G(x)=\int d^4x^{\prime }{\frac{{\delta G(x)}}{{\delta g(x^{\prime })}}}
\Lie g(x^{\prime })\quad  \label{Lie4}
\end{equation}
which follows from (\ref{Lie3}), in condensed notation takes the form 
\begin{equation}
\Lie G=G_{,a}(\Lie g)^a \,\, , 
\label{f9}
\end{equation}
where in addition we omitted all ``irrelevant'' indexes.

\section{Back Reaction Problem for Cosmological Perturbations}

We consider a homogeneous, isotropic Universe with small perturbations.
This
means we can find a coordinate system $(t,x^i)$ in which the metric $(g_{\mu
\nu })$ and matter $(\varphi )$ fields, denoted for brevity by the collective
variable $q^a$, can be written as 
\begin{equation}
q^a = q_0^a + \delta q^a \, ,  \label{eq:8}
\end{equation}
where the background field $q_0^a$ is  defined as  homogeneous part of $q^a$
on the hypersurfaces of constant time $t$ and therefore $q_0^a$ depends only
on the time variable $t$ (we recall that the variables $t$ and $x^i$ are
included in the index $a$) . The perturbations $\delta q^a$ depend on both
time and spatial coordinates, and by assumption they are small: 

\begin{equation} | \delta q^a | \, \ll \, q_0^a \,\, .
\end{equation}
{}From our definition of the background component  $q_0^a$ it follows  that
the spatial average of $\delta q^a$ vanishes: 

\begin{equation}
\langle \delta q^a \rangle= \lim_{V\to \infty } \,\, {\frac{\int_V\delta
q^ad^3x}{\int_Vd^3x}} = 0 \,\, .
\end{equation}
Spatial averaging is defined with respect to the background metric, not
with respect to the perturbed metric as was done in Ref. \cite{Seljak}. 
Our definition is the appropriate one when establishing what 
``perturbations" are and when constructing the general 
back reaction framework.  The averaging of Ref. \cite{Seljak} is
appropriate when discussing the ``expected" values of physical quantities
for real observers in systems with fluctuations on scales 
smaller than the Hubble radius.

The Einstein equations 
\begin{equation}
G_{\mu \nu }\,-\,8\pi GT_{\mu \nu }:=\Pi _{\mu \nu }=0  \label{Einstein}
\end{equation}
can be expanded in a functional power series in $\delta q^a$ about the
background $q_0^a$, if we treat $G_{\mu \nu }$ and $T_{\mu \nu }$ as
functionals of $q^a$

\begin{equation}
\Pi (q_0^a) + \left. \Pi _{,a} \right|_{q_0^a} \delta q^a + 
{\frac 12} \left. \Pi _{,ab} \right|_{q_0^a} \delta q^a\delta q^b +
{\cal {O}}(\delta q^3) = 0  \,\, .
\label{EE2}
\end{equation}

To lowest order the background $q_0^a$ should obey the Einstein
equations
\begin{equation}
\Pi (q_0^a)=0  \label{eq1}
\end{equation}
and  the fluctuations $\delta q^a$ satisfy the linearized Einstein equations

\begin{equation}
\quad \Pi _{,a}(q_0)\delta q^a=0 \,\,.
\label{E_Lin}
\end{equation}
By definition, the spatial average of the linear term in $\delta q^a$ in 
(\ref{EE2}) vanishes. The term quadratic in $\delta q^a$, however, does
not.
Therefore the spatial averaging of (\ref{EE2}) leads to higher order
corrections in the equations describing the behaviour of the homogeneous
background mode. Thus the ``corrected'' equations which take into account
the back reaction of small perturbations on the evolution of the
background are

\begin{equation}  \label{BR1}
\Pi (q_0^a ) = \thinspace - {\frac 12} \langle \Pi _{,ab}\delta
q^a\delta q^b \rangle \thinspace
\end{equation}

At first sight, it seems natural to identify the quantity on the right hand
side of (\ref{BR1}) as the effective EMT of perturbations which describes
the back reaction of fluctuations on the homogeneous background. However, it
is not a gauge invariant expression and, for instance, does not vanish for
``metric perturbations" induced by a coordinate transformation in Minkowski
space-time.

In the next section we will rewrite the back-reaction equations in a
manifestly gauge-invariant form. First, however, we want to show that the
physical content of (\ref{BR1}) is independent of the gauge chosen to do the
calculation in, provided that we take into account that the background
variables change to second order under a gauge transformation.

The coordinate transformation (\ref{Lie1}) induces (to second order in
perturbation variables) the following diffeomorphism transformation of a
variable $q$ 

\begin{eqnarray}
q &=&q_0+\delta q\rightarrow e^{-{\cal L}_\xi }\left( q_0+\delta q\right)  
\nonumber \\
&=&q_0+\delta q-{\cal L}_\xi q_0-{\cal L}_\xi \,\delta q+{\frac 12}\,{\cal L}
_\xi ^2q_0
\end{eqnarray}
Hence, to linear order, the change in $\delta q$ is 

\begin{equation}
\delta q \rightarrow \delta \tq = \delta q-{\cal L}_\xi q_0\,,
\label{TR1}
\end{equation}
while, to second order, the background variable transforms nontrivially as 
\begin{equation}
q_0\rightarrow \tilde q_0= q_0 - \langle{\cal L}_\xi \delta q \rangle
 + {\frac 12} \langle {\cal L}_\xi ^2 q_0 \rangle \ ,  
\label{TR2}
\end{equation}
where $\langle \xi \rangle=0$ has been assumed.

In order to prove that the equations (\ref{BR1}) are independent of
gauge, we must show that 

\begin{equation}
\Pi (q_0) = - {\frac 12} \langle \Pi _{,ab} \delta q^a \, \delta q^b \rangle 
\,\,\Leftrightarrow
\,\,\Pi \left( \tilde q_0\right) =-{\frac 12} \langle\Pi _{,ab} \delta
\tilde q
^a\delta \tilde q^b\rangle  \label{TR3}
\end{equation}
(to second order in perturbation variables). Making use of (\ref{TR1}) and 
(\ref{TR2}) we obtain 

\begin{equation}
\Pi \left( \tilde q_0\right) =\Pi (q_0)-\Pi _{,a} \langle {\cal L}_\xi
\delta q^a\rangle +
{\frac 12}\Pi _{,a} \langle {\cal L}_\xi ^2q_0^a\rangle  \label{TR4}
\end{equation}
and 

\begin{eqnarray}
-{\frac 12} \langle \Pi _{,ab}\,  \delta \tilde q^a\delta \tilde
q^b\,\rangle= &-& {\frac 12}
\langle \Pi _{,ab}\, \delta q^a\delta q^b\rangle  \label{TR5} \\
&+& \langle \Pi _{,ab}\, {\cal L}_\xi q_0^a\delta q^b\rangle-{\frac 12}
\langle \Pi_{,ab}\, {\cal L}
_\xi q_0^a{\cal L}_\xi q_0^b\rangle\,.  \nonumber
\end{eqnarray}
In order to show that the extra terms on the right hand sides of (\ref{TR4})
and (\ref{TR5}) cancel out in (\ref{TR3}), we make use of (\ref{Lie4}) 
in the equation 

\begin{equation}
 \langle (e^{-{\cal L}_\xi } - 1) \Pi (q_0 + \delta q) \rangle \, = \, 0 \, . 
\end{equation}
Expanding this equation to second order in perturbations, we obtain the 
identity
 
\begin{equation}
{\frac 12} \langle \Pi _{,a} {\cal L}_\xi ^2q_0^a\rangle 
+ {\frac 12} \langle \Pi _{,ab} {\cal L}_\xi q_0^a{\cal L}_\xi q_0^b\rangle
- \langle \Pi _{,a} {\cal L}_\xi \delta q^a \rangle 
- \langle \Pi _{,ab} {\cal L}_\xi q_0^a \delta q^b \rangle \, = \, 0 \label{TR6}
\end{equation}
which completes the proof that the extra terms mentioned above cancel out.

\vskip .20in

\section{Gauge Invariant Form of the Back Reaction Equations}

Although we have shown in the previous section that the physical content of
Eqs. (\ref{BR1}) should be the same  in all coordinate systems, it is
useful for many purposes to recast these equations in an explicitly
gauge-invariant way. In particular, this will allow us to define a gauge
invariant EMT for cosmological perturbations.

We start by writing down the metric for a perturbed spatially flat FRW
Universe 
\begin{eqnarray}
ds^2 = & &  (1+2 \phi) dt^2 -2 a(t) (B_{,i} - S_i ) dx^i dt
\label{PV1} \\
& - & a^2(t) [ (1-2\psi) \delta_{ij} + 2 E_{,ij} + F_{i,j} + F_{j,i} +
h_{ij} ] dx^i dx^j \, ,  \nonumber
\end{eqnarray}
where $a(t)$ is the scale factor, and where the 3-scalars $\phi , \psi,
B$ and $E$ characterize scalar metric
perturbations. The symbols $S_i$ and $F_i$ are transverse 3-vectors
(giving vector fluctuations), and $h_{ij}$ is a traceless transverse
3-tensor (gravity
waves).

We will now attempt to construct gauge-invariant variables $Q$ from the
gauge-dependent quantities $q$. To do that let us take  the set of 4
quantities $X^\mu $ (not necessarily a four-vector) 
and form a Lie operator
with $X^\mu$ (denoted by ${\cal L}_X$), treating $X^\mu $ formally as a
four-vector. 
Later on, after we specify the properties which $X^\mu$ should
satisfy if we want it to help build gauge
invariant variables, we will then construct $X^\mu$ explicitly out of the
metric perturbation variables. However, let us leave $X^\mu$ unspecified
for a moment. Using 
${\cal L}_X$ we define the new variable $Q$ according to the prescription  
\begin{equation}
Q=e^{{\cal L}_X}q \,\, .  
\label{GV1}
\end{equation}
If we attempt the same construction after performing a gauge transformation
 (with parameter $\xi $),
then taking into account that as a result of this transformation 
$X\rightarrow {\tilde X}$ and 

\begin{equation}
q\rightarrow {\tilde q}=e^{-{\Lie}}q \,\, ,
\end{equation}
we obtain 

\begin{equation}
{\tilde Q}(x)=e^{{\cal L}_{\tilde X}}e^{-{\Lie}}q(x) \,\, .  
\label{GV2}
\end{equation}
If we demand that $Q$ is gauge-invariant, that is ${\tilde Q}=Q$, then
comparing (\ref{GV1}) and (\ref{GV2}) we arrive at the
condition 
\begin{equation}
e^{{\cal L}_{\tilde X}}=e^{{\cal L}_X}e^{\Lie}\,.  \label{GV3}
\end{equation}
which imposes strong restrictions on the transformation law  
$X\rightarrow {\tilde X}$.
{}From (\ref{GV3}) and by using the
Baker-Campbell-Hausdorff formula for the products of exponentials of
operators, one can easily find that the condition for gauge-invariance of $Q$
implies that under a diffeomorphism generated by $\xi $,
 
\begin{equation}
X^\mu \rightarrow {\tilde X}^\mu =X^\mu +\xi ^\mu +{\frac 12}[X,\xi ]^\mu
 + \cdots \,,
\end{equation}
where $\cdots$ denote terms of cubic and higher order.

Note that $Q$ is a gauge invariant variable characterizing both the
background 
\begin{equation}
Q_0^a = q_0^a + \langle {\cal L}_X\delta q^a\rangle +\frac 12
\langle {\cal L}_X^2q_0^a\rangle
\end{equation}
and the linearized perturbations 
\begin{equation}  \label{GV4}
\delta Q^a= \delta q^a+ {\cal L}_Xq_0^a.
\end{equation}

Making use of the gauge invariant variable $Q$ we can recast the
back-reaction problem (\ref{BR1}) in a manifestly gauge invariant form. If $q
$ satisfies the Einstein equations, it follows from our basic identity (\ref
{Lie3}) that 
\begin{equation}
e^{{\cal L}_X}\Pi (q)=\Pi (e^{{\cal L}_X}q)=\Pi (Q)=0.
\end{equation}
Expanding the above equation to second order in $\delta Q$ and taking the
spatial average of the result yields 
\begin{equation}
\Pi (Q_0)=-\frac 12 \langle \Pi _{,ab} \delta Q^a\delta Q^b\rangle 
= - \frac 12
\langle G_{,ab}\delta Q^a\delta Q^b \rangle +4\pi G \langle T_{,ab}\delta
Q^a\delta Q^b\rangle ,  \label{BR4}
\end{equation}
which is the desired gauge invariant form of the back-reaction equation.
Reinserting tensor indices, the above equation can be rewritten as 
\begin{equation}
G_{\mu \nu }(Q_0)=8\pi G[T_{\mu \nu }(Q_0)+\tau _{\mu \nu }(\delta Q)],
\label{BR2}
\end{equation}
where 

\begin{equation}  \label{BR3}
\tau _{\mu \nu }(\delta Q)\equiv -\frac 1{16\pi G}
\langle \Pi_{\mu\nu,ab}\delta Q^a\delta Q^b \rangle   
\label{teff}
\end{equation}
can be interpreted as the gauge-invariant effective EMT for cosmological
perturbations.

At this point, however, we must return to the question of what $X^{\mu}$ is.
It is a question of linear algebra to find the linear combinations of the
perturbation variables of (\ref{PV1}) that have (to linear order) the
required transformation properties, namely 
\begin{equation}
X^{\mu} \rightarrow X^{\mu} + \xi^{\mu} \, .
\end{equation}
The solution we will use is 
\begin{equation}  \label{aux1}
X^{\mu} = [a(B - a {\dot E}) \, , \, -E_{,i} -
F_i] \, ,
\end{equation}
where a dot denotes a derivative with respect to the time variable $t$.
But this choice is not unique. There is a four parameter family of
possible choices labelled by real parameters 
$\alpha, {\bar \alpha},  \gamma$  and $\sigma$. The component 
$X^0$ is 

\begin{equation}
X^0 = \alpha a(B - a {\dot E}) + (1 - \alpha)
({\frac{{a} }{{\dot a}}} \psi)  \, ,
\end{equation}
the $X^i$ components have a traceless piece

\begin{equation}
X^i_{Tr} = - \sigma F_i + (1 - \sigma) \int^t_0 dt^{\prime}
{1 \over a} S_i(t^{\prime})
\end{equation}
and a trace 

\begin{equation}
X^i_{L} = X_{L,i} \, ,
\end{equation}
with the function $X_L$ defined as
\begin{equation}  \label{aux2}
X_L = - \gamma E + (1 - \gamma) \int^t_0 dt^{\prime} {1 \over a} \left[
{1 \over a} X^0({\bar \alpha}; t^{\prime}) - 
B(t^{\prime}) \right] \,\, .
\end{equation}
Finally, 
\begin{equation}
X^i = X^i_{Tr} + X^i_L \,\, .
\end{equation}
Demanding regularity in the limit where the expansion rate vanishes
forces  $\alpha = {\bar \alpha} = 1$. 
In this case, the dependence on $\gamma$ drops out of
Eq. (\ref{aux2}) and we are left with a one-parameter degeneracy of $X^{\mu}$
labelled by $\sigma$.

\vskip .20in
\section{Energy-Momentum Tensor for Scalar Perturbations}

In this and the following section, we calculate the effective EMT for scalar
and tensor perturbations, respectively. Since vector modes decay in an
expanding Universe, we shall in the following take them to be absent.

In models such as the inflationary Universe scenario, scalar and tensor
modes are statistically independent Gaussian random fields. In this case,
the effective EMT (4.10) separates into two independent pieces, the first
due to the scalar perturbations, the second due to the tensor modes. 
\begin{equation}
\tau _{\mu \nu }(\delta Q)=\tau _{\mu \nu }^{scalar}(\delta Q)+\tau _{\mu
\nu }^{tensor}(\delta Q)\,
\end{equation}

Note that if we neglect vector perturbations then it is enough to consider
only scalar modes in $X^\mu $. Hence, the contribution to the effective
EMT (\ref{BR3}) coming from the term ${\cal L}_Xq_0^a$ appearing in
$\delta Q^a$
is a contribution to $\tau _{\mu \nu }^{scalar}$ alone.

It is easy to verify that for our choice of $X^\mu $, namely, 
\begin{equation}
X^\mu = [a(B - a {\dot E}),-E,_i] \, ,
\end{equation}
the variables $\delta Q^a$ of (\ref{GV4})
for the metric perturbations correspond to Bardeen's gauge invariant
variables$^{\cite{Bardeen}}$. In fact, the application of (\ref{GV4})
yields the following gauge invariant metric tensor: 
\begin{equation}
\delta g_{\mu \nu }^{(gi)}=\delta g_{\mu \nu }+{\cal L}_Xg_{\mu \nu }^{(0)} \, ,
\end{equation}
and from the time-time and diagonal spatial components, respectively, we
immediately obtain the two gauge invariant variables, 
\[
\phi ^{(gi)}\equiv \Phi = \phi + [a (B - a {\dot E})]^{\cdot}, 
\]
\begin{equation}
\psi ^{(gi)}\equiv \Psi = \psi - {\dot a} (B - a {\dot E}).
\end{equation}
For the remaining components of the metric tensor, the gauge invariant
combinations vanish. 

Hence, calculating the general gauge-invariant effective EMT (\ref
{BR3}) reduces to calculating 
\begin{equation}
\tau _{\mu \nu }=-{\frac 1{16\pi G}}\,<\Pi _{\mu \nu ,ab}\,\delta q^a\delta
q^b>
\end{equation}
in longitudinal gauge ($B=E=0)$, in which 
\begin{equation}
\label{longit}
ds^2=(1+2\phi )\,dt^2-a^2(t)\,(1-2\psi )\,\delta _{ij}dx^i\,dx^j
\end{equation}

For many types of matter (scalar fields included), $T_{ij}$ is diagonal to
linear order in $\delta q$. In this case, it follows from the linearized
Einstein equations that 
\begin{equation}
\phi =\psi 
\end{equation}
Thus, in longitudinal gauge the variable $\phi $ entirely characterizes the
metric perturbations. We shall consider scalar field matter, in which case
the linear matter fluctuations are described by $\delta \varphi $. The
motivation for our choice of matter follows since we have applications of
our formalism to inflationary cosmology in mind. The linearized Einstein
equations also relate $\delta \varphi $ and $\phi $. In fact, there is a
single gauge invariant variable characterizing linearized scalar
fluctuations. 

We will now calculate $\tau _{\mu \nu }^{(2)}$ for scalar perturbations
(metric (\ref{longit})). The contribution of gravitational part to the
effective EMT
can be easily calculated with the help of the formulae (see, e.g., in
\cite{MTW}
pg.965, Eq. (35.5g) a\& b): 
\begin{eqnarray}
R_{\mu \nu }^{(2)}={\frac 12}R_{\mu \nu ,ab}\,\delta g^a\,\delta g^b 
&=&{\frac 12}\left[ {\frac 12}\,\delta g_{|\mu }^{\alpha \beta }\,\delta
g_{\alpha \beta |\nu } + 
\delta g^{\alpha \beta }\left( \delta g_{\alpha\beta |\mu \nu }
+\delta g_{\mu \nu |\alpha \beta }-\delta g_{\alpha \mu |\nu \beta
}-\delta g_{\alpha \nu |\mu \beta }\right) \right.  \nonumber \\
&& + \delta g_\nu ^{\alpha |\beta }\left( \delta g_{\alpha \mu |\beta}
-\delta g_{\beta \mu |\alpha }\right) -  \label{var} \\
&&-\left. \left( \delta g_{|\beta }^{\alpha \beta }-{\frac 12}\,\delta
g^{|\alpha }\right) \left( \delta g_{\alpha \mu ;\nu }+
\delta g_{\alpha\nu|\mu }-
\delta g_{\mu \nu |\alpha }\right) \right] \,.
\nonumber
\end{eqnarray}
if we substitute in these expressions the metric (\ref{longit}).
In this formula $|$ denotes covariant derivatives with respect to the
background metric.
Expanding the energy-momentum tensor
for a scalar field 
\begin{equation}
T_{\mu \nu }=\varphi _{,\mu }\varphi _{,\nu }-g_{\mu \nu }\left[ {\frac 12}
\varphi ^{,\alpha }\varphi _{,\alpha }-V(\varphi )\right] 
\end{equation}
to second order in $\delta \varphi $ and $\delta g$ and combining it with
the result for $G_{\mu \nu }^{(2)}$ we obtain from (\ref{BR3})

\begin{eqnarray}  \label{tzero}
\tau_{0 0} &=& \frac{1}{8 \pi G} \left[ + 12 H \langle \phi \dot{\phi} \rangle
- 3 \langle (\dot{\phi})^2 \rangle + 9 a^{-2} \langle (\nabla \phi)^2
\rangle \right]  \nonumber \\
&+& \meio \langle ({\delta\dot{\varphi}})^2 \rangle + \meio a^{-2} \langle
(\nabla\delta\varphi)^2 \rangle  \nonumber \\
&+& \meio V''(\varphi_0) \langle \delta\varphi^2 \rangle + 2
V'(\varphi_0) \langle \phi \delta\varphi \rangle \quad ,
\end{eqnarray}
and 
\begin{eqnarray}  \label{tij}
\tau_{i j} &=& a^2 \delta_{ij} \left\{ \frac{1}{8 \pi G} \left[ (24 H^2 + 16 
\dot{H}) \langle \phi^2 \rangle + 24 H \langle \dot{\phi}\phi \rangle
\right. \right.  \nonumber \\
&+& \left. \langle (\dot{\phi})^2 \rangle + 4 \langle \phi\ddot{\phi}\rangle
- \frac{4}{3} a^{-2}\langle (\nabla\phi)^2 \rangle \right] + 
4 \dot{{\varphi_0}}^2 \langle \phi^2 \rangle  \nonumber \\
&+& \meio \langle ({\delta\dot{\varphi}})^2 \rangle 
- {1 \over 6} a^{-2} \langle(\nabla\delta\varphi)^2 \rangle 
- 4 \dot{\varphi_0} \langle \delta \dot{\varphi}\phi \rangle  \nonumber \\
&-& \left. \meio \, V''(\varphi_0) \langle \delta\varphi^2
\rangle + 2 V'( \varphi_0 ) \langle \phi \delta\varphi \rangle
\right\} \quad ,
\end{eqnarray}
where $H$ is the Hubble expansion rate, and where we have used the fact that 
$\phi = \psi$ for theories in which $\delta
T_{ij}$ is diagonal at linear order.

Before discussing the long and short wavelength limits of the equation of
state satisfied by the cosmological perturbations, let us briefly recall a
few crucial points from the theory of linear fluctuations (see Ref.
\cite{MFB92} for a comprehensive overview).

The simplest way to derive the equations of motion satisfied by the gauge
invariant perturbation variables is to go to longitudinal gauge ($B=E=0$),
in which the gauge invariant variables $\Phi $ and $\Psi $ coincide with the
metric fluctuations $\phi $ and $\psi $. The equations of motion 
(\ref{E_Lin})

\begin{equation}
\delta G_\nu ^\mu =8\pi G\delta T_\nu ^\mu 
\end{equation}
derived in this gauge coincide with the equations for Bardeen's gauge
invariant variables.

{}For scalar field matter, as mentioned above, $\phi =\psi$. In this case,
the 00 and $ii$ perturbation equations combine into a second
order differential equation for $\Phi$ which on scales larger than the
Hubble radius and for a time-independent background equation of state has
the solution
\be
\label{phi} \phi ({\bf x},t) \simeq c({\bf x})
\ee
(modulo the decaying mode). The $0i$
equations give a constraint relating $\phi $ and $\delta \varphi$: 
\begin{equation}
\dot \phi +H\phi = 4\pi G\dot \varphi _0\,\delta \varphi \,.  \label{constr}
\end{equation}

Our aim is to work out $\tau _{\mu \nu }$ in an inflationary Universe. In
most models of inflation, exponential expansion of the Universe results
because $\varphi _0$ is rolling slowly, i.e. 
\begin{equation}
\dot \varphi _0\simeq -{\frac{V^{\prime }}{3H}}\,,  \label{eom}
\end{equation}
where a prime denotes the derivative with respect to the scalar matter
field. The $\dot \phi $ term in Eq.(\ref{constr}) for the nondecaying
mode of perturbations is
proportional to a small slow roll parameter and therefore can be
neglected. Thus from Eqs. (\ref{constr}) and (\ref{eom}) we obtain 
\begin{equation}
\delta \varphi =-{\frac{2V}{V^{\prime }}}\,\phi \,.  
\label{constr2}
\end{equation}

When considering the contributions of long wavelength fluctuations to $\tau
_{\mu \nu }$, we can neglect all terms in (\ref{tzero}) and (\ref{tij})
containing gradients and $\dot \phi $ factors. Because of the ``slow
rolling'' condition (\ref{eom}), the terms proportional to 
$\dot \varphi_0^2$
and $\dot H$ are negligible during inflation (but they become important at
the end of inflation). Hence, in this approximation 

\begin{equation}
\tau _{00}\simeq {\frac 12}\,V^{\prime \prime }\,<\delta \varphi
^2>+2V^{\prime }<\phi \delta \varphi >
\end{equation}
and 

\begin{equation}
\tau _{ij}\simeq a^2\delta _{ij}\,\left\{ {\frac 3{\pi G}}H^2<\phi ^2>
-{\frac 12}\,V^{\prime \prime }<\delta \varphi^2 >+2V^{\prime }<\phi \delta
\varphi >\right\} \,.
\end{equation}
Making use of (\ref{constr2}), this yields 

\begin{equation}
\rho _s\equiv \tau _0^0\cong \left( 2\,{\frac{{V^{\prime \prime }V^2}}
{{V^{\prime }{}^2}}}-4V\right) <\phi ^2>  \label{tzerolong}
\end{equation}
and 
\begin{equation}
p_s\equiv -\frac 13\tau _i^i\cong -\rho ^{(2)}\,.  \label{tijlong}
\end{equation}

Thus, we have shown that the long wavelength perturbations in an
inflationary Universe have the same equation of state $p_s=-\rho _s$ as the
background.

One of the main results which emerges from our analysis is that 
\begin{equation}
\rho _s<0
\end{equation}
for the long wavelength cosmological perturbations in all realistic
inflationary models. Thus, the effective $\tau _{\mu \nu }$ counteracts the
cosmological constant driving inflation. Note that the same sign of $\rho $
emerges when considering the vacuum state EMT of a scalar field in a fixed
background de Sitter space-time (see e.g. Refs. \cite{Ford,Mottola}).

For short wavelength fluctuations ($k \gg aH)$, both $\phi $ and $\delta
\varphi $ oscillate with a frequency $\propto k$. In this case
(see (\ref{constr})): 
\begin{equation}
\phi \sim 4\pi G\,{\frac{ia}k}\,\dot \varphi _0\delta \varphi \,.
\end{equation}
Hence, it follows by inspection that all terms containing $\phi $ in (\ref
{tzero}) and (\ref{tij}) are suppressed by powers of $Ha/k$ compared to the
terms without dependence on $\phi $, and therefore 
\begin{equation}
\tau _{00}\simeq {\frac 12}<(\delta \dot \varphi )^2>+{\frac 12}
a^{-2}<(\nabla \delta \varphi )^2>+{\frac 12}\,V^{\prime \prime }(\varphi
_0)<\delta \varphi ^2>
\end{equation}
and 
\begin{equation}
\tau _{ij}=a^2\delta _{ij}\,\left\{ {\frac 12}<(\delta \dot \varphi )^2>
-{\frac 16}a^{-2}<(\nabla \delta \varphi )^2>-{\frac 12}\,V^{\prime \prime
}(\varphi _0)<\delta \varphi ^2>\right\}
\end{equation}
which is a familiar result.

\section{Energy-Momentum Tensor for Gravitational Waves}

\vskip .10in

In the case of gravitational waves the metric takes the form 
\begin{equation}
ds^2=dt^2-a^2(t)(\delta _{ik}+h_{ik})dx^idx^k
\end{equation}
where $h_{ik}$ is defined as the transverse traceless part of the metric
perturbations, and therefore the components $\delta g_{ik}\equiv $ $h_{ij}$
are gauge invariant themselves.

In the absence of matter fluctuations, the effective EMT \thinspace $\tau
_{\mu \nu }$ is given by the first term in (\ref{BR4}). Making use of the
second variation of $R_{\mu \nu }$ given by (\ref{var}) and the equation of
motion for gravitational waves 
\begin{equation}
\stackrel{..}{h}_{ij}+3{\frac{\stackrel{.}{a}}a}\stackrel{.}{h}_{ij}
-\frac 1{a^2}\nabla ^2h_{ij}=0 \, , \label{geom}
\end{equation}
we obtain 
\begin{equation}
8 \pi G \tau _{00}=
\, {\frac{\stackrel{.}{a}}a}<\stackrel{.}{h}_{k\ell}h_{k\ell }>
+ {\frac 18}\left( <\stackrel{.}{h}_{k\ell }\stackrel{.}{h}_{k\ell}> 
+ \frac 1{a^2}<h_{k\ell ,m}h_{k\ell ,m}>\right)  \label{tgzero}
\end{equation}
and 
\begin{eqnarray}
8 \pi G \tau _{ij} 
&=&\delta _{ij}a^2\left\{ {\frac 3{8a^2}}<h_{k\ell,m}h_{k\ell
,m}>-{\frac 38}<\stackrel{.}{h}_{k\ell }\stackrel{.}{h}_{k\ell }>\right\} 
\nonumber \\
+\frac 12a^2 &<&\stackrel{.}{h}_{ik}\stackrel{.}{h}_{kj}>+\frac 14<h_{k\ell
,i}h_{k\ell ,j}>-{\frac 12}<h_{ik,\ell }h_{jk,\ell }>\,.  \label{tgij}
\end{eqnarray}
The first expression can be interpreted as the effective energy density of
gravitational waves 
\begin{equation}
\rho _{gw}=\tau _0^0 \,\, .
\end{equation}

The relation between $\tau_{ij}$ and the quantity which we could
naturally interpret as an effective pressure is not so straightforward as it
looks at first glance. The problem is that the energy momentum tensor for
the gravity waves $\tau _{\mu \nu }$ is not conserved itself, that is,  
$\tau _{\nu |\mu }^\mu \neq 0.$
This is
not surprising since the gravitational perturbations ``interact'' with
the background and only the total EMT must be conserved, 

\begin{equation}
\label{conserv}
(T_\nu ^\mu (Q_0)+\tau _\nu ^\mu (\delta Q))_{|\mu }=0
\end{equation}
as a consequence of Bianchi identities for background, $G_\nu ^\mu
(Q_0)_{|\mu }=0.$ In fact expanding the exact conservation law 
$T_{\nu ;\mu}^\mu =0$ in perturbations to second order and averaging
the resulting equation we obtain 

\begin{equation}
T_\alpha ^\beta (Q_0)_{|\beta } = 
	- \langle ^{(2)} \Gamma_{\beta \gamma}^\beta \rangle \,
 T_\alpha^\gamma  + 
	\langle ^{(2)} \Gamma _{\alpha \beta }^\gamma \rangle \,
 T_\gamma^\beta  
\label{cons}
\end{equation}
Deriving Eq. (\ref{cons}) we took into account that in our case (when we
have only gravity waves)  matter perturbations are absent. The energy
momentum
tensor for the background is diagonal and isotropic, that is, 
$ T_0^0=\rho^{(0)} $ and $ T_k^i=-p^{(0)} \delta_k^i$. Also, we will consider
only isotropic fields of gravitational waves. In such a case the
conservation law (\ref{conserv}), taking into account (\ref{cons}), can
be written down explicitly in a familiar form as 
\begin{equation}
\dot{\rho }_{gw}+3H(\rho_{gw}+p_{gw})=0 
\end{equation}
where 
\begin{equation}
p_{gw}=-\frac 13\tau _i^i-\frac 1{3H} \langle^{(2)}\Gamma_{\beta 0}^\beta
\rangle 
(\rho^{(0)} + p^{(0)})  \label{pressure}
\end{equation}
can be interpreted as the pressure of gravitational waves 

The second term in (\ref{pressure}) does not contribute to the pressure only in
a de Sitter Universe in which $p^{(0)}=-\rho^{(0)}$. In this case the
term 
$-\frac 13\tau _i^i$ 
itself can be interpreted as a pressure. A similar but
a bit more complicated analysis can be done for the scalar perturbations,
for which one third of the trace of the spatial part of the EMT can be
interpreted as a pressure also only in a de Sitter Universe.

As for scalar perturbations, we will now study the equation of state for
gravity waves both in the short and long wavelength limits in various models
for the evolution of the Universe. First, assuming that the field of 
gravity waves is isotropic and by averaging the diagonal elements of
(\ref{tgij}) we
obtain the contribution of $\tau $ to the pressure 
\begin{equation}
-{{8 \pi G} \over {3}}\,\tau _i^i={\frac 7{24a^2}}\,<h_{k\ell ,m}h_{k\ell
,m}>-{\frac 5{
24}}<\stackrel{.}{h_{k\ell }}\stackrel{.}{h_{k\ell }}>\,.  \label{cont}
\end{equation}
For fluctuations with wavelength smaller than the Hubble radius, the first
term on the right hand side of (\ref{tgzero}) and the second term in (\ref
{pressure}) are negligible and the time average of the temporal and spatial
gradient terms are the same. Hence 
\begin{equation}
p_{gw}={\frac 13}\rho _{gw} = {{1} \over {8 \pi G}} 
{\frac 1{12a^2}}\ll h_{k\ell ,m}h_{k\ell ,m}\gg
\end{equation}
where $\ll \,\gg $ indicates that in addition to spatial average, a time
average over a period $T \ll H^{-1}$ has been taken. As expected, short
wavelength gravitational waves behave
like radiation, independent of the evolution of the background, and their
energy density decays as $a^{-4}(t)$.

In the case of long wavelength gravitational waves the calculations are less
straightforward. First, we consider a de Sitter background : 
\begin{equation}
a(t)=e^{H (t-t_0)}
\end{equation}
where $t_0$ is a reference time. Let us take an isotropic field of gravity
waves with comoving wavenumbers $k$. 
For the nondecaying mode of long
wavelength gravity waves ($k/a\ll H)$, the solution of Equation (\ref{geom}) is 

\begin{equation}
h_{ij}= A_k \epsilon _{ij}\left[ 1 + \frac 12 \left(\frac{k}{aH}\right)^2 
+O((\frac{k}{aH})^3) \right] e^{i\vec{k}\vec{x}} \,,  
\label{hij}
\end{equation}
where $A_k$ is a constant (related to the spectrum of gravity waves) and
$\epsilon _{ij}$ is the polarization
tensor. Substituting this solution in Formula (\ref{tgzero}) we obtain the
following expression for the energy density of the gravity waves to lowest
order in $k$ : 

\begin{equation}
\rho _{gw} \simeq -{{1} \over {8 \pi G}}{\frac 78} \frac{k^2}{a^2}
	\langle |A_k|^2 \epsilon_{ij} \epsilon^{ij} \rangle
\label{rho2}
\end{equation}
and correspondingly from (\ref{cont}) we derive that the pressure is: 
\begin{equation}
p_{gw}\simeq {-}\frac 13\,\rho _{gw}.
\end{equation}
Note that the whole contribution to the pressure in this case comes from the
first term in (\ref{pressure}) . The energy density of long wavelength 
gravity waves decays as $\sim a^{-2}$.

In a radiation dominated Universe the scale factor increases as 
\begin{equation}
a(t)=\left( \frac{t}{t_0} \right)^{1/2} \,\, ,
\end{equation}
and the solution for long wavelength gravity waves ($k \ll Ha)$ is 
\[
h_{ij}= A_k \epsilon _{ij}\left[ 1 - \frac {1}{6} \left(\frac{k}{aH}\right)^2 
+O((\frac{k}{aH})^3) \right] e^{i\vec{k}\vec{x}} \,,  
\]
Inserting these results into (\ref{tgzero}) and (\ref{tgij}) yields 
\begin{equation}
\rho _{gw}\simeq - 
\frac{1}{8 \pi G} \frac {5}{24} \, \frac{k^2}{a^2}  
	\langle |A_k|^2 \epsilon _{ij} \epsilon^{ij} \rangle
\label{eos1}
\end{equation}
and 
\begin{equation}
p_{gw}\simeq {-}\frac 13\,\rho _{gw}\,.  \label{eos2}
\end{equation}
In this case both of the terms in (\ref{pressure}) give comparable 
contributions to the pressure, namely the contribution of the first term 
there is $p_1 = \frac{21}5p_{gw}$ and the contribution of the second
term is negative and equal to $p_2=-\frac{16}5p_{gw}.$ As in the previous
case the energy density decays as $a^{-2}.$

\section{Back-Reaction in Inflationary Universe}

\vskip0.1in

As an application of the formalism developed in this paper, we will study
the effects of back-reaction in inflationary cosmology. To be specific, we
consider a single field chaotic inflation model$^{\cite{Linde}}$ and take
the inflaton potential to be 
\begin{equation}
V(\varphi )={\frac 12}m^2\varphi ^2\,.  \label{pot}
\end{equation}
Furthermore, we specify an initial state at a time $t_i$ in which the
homogeneous inflaton field has the value $\varphi _0(t_i)$ and  the
fluctuations are minimal.

We will focus on the contribution of long wavelength modes to the effective
EMT, which by assumption vanishes at the initial time $t_i$. Provided that 
$\varphi _0(t_i)$ is sufficiently large, the slow-rolling conditions are
satisfied and exponential expansion will commence. At this point, quantum
vacuum fluctuations begin to generate perturbations on scales which
``leave'' the Hubble radius (i.e. whose physical wavelength becomes larger
than the Hubble radius). As time proceeds, more modes leave the Hubble
radius, and hence the contribution to the effective EMT builds up. We wish
to estimate the magnitude of the resulting effective EMT.

As discussed in Setion 5, scalar metric perturbations in this model are
characterized by a single function $\phi_k$ (where $k$ stands for the
comoving wave number). From the constraint equation (\ref{constr2}) it
follows that 
\begin{equation}
{\delta\varphi}_k \simeq - \varphi_0 \phi_k \, .
\end{equation}
Hence, the dominant terms in the effective energy-momentum tensor $\taumn$
(see (\ref{tzerolong}) and (\ref{tijlong})) are proportional to the
correlator 
$\langle \phi^2 \rangle$. The amplitudes of $\phi_k$ are known from the
theory of linear cosmological perturbations. Using the results for $\phi_k$
valid during inflation$^{\cite{MFB92}}$ we obtain 
\begin{equation}  \label{correl}
\langle \phi^2 (t) \rangle \, = \, \int_{k_i}^{k_t} \frac{dk}{k} |\phi_k|^2
\, = \, \frac{m^2 M_P^2}{32 \pi^4 \varphi_0^4 (t)} \int_{k_i}^{k_t} 
\frac{dk}{k} {\left[ \ln{\ \frac{H(t) a(t)}{k} } \right]}^2 \, ,
\end{equation}
where $t$ denotes physical time, and $M_P$ is the Planck mass. The integral
runs over all modes with scales larger than the Hubble radius, i.e. 
\begin{equation}
k \, < \, k_f (t) \, = \, H(t) a(t) \, ,
\end{equation}
but smaller than the Hubble radius at the initial time $t_i$, i.e. 
\begin{equation}
k \, > \, k_i \, = \, H(t_i) a(t_i) \, .
\end{equation}
The infrared cutoff $k_i$ is a consequence of our choice of initial state.

For the potential $V(\varphi)$ considered here, the scale factor is given by 
\begin{equation}
a(t) \, = \, a(t_i) exp({\frac{{2 \pi} }{{M_P^2}}} [\varphi_0^2(t_i) -
\varphi_0^2(t)]) \, .
\end{equation}
The intregral over $k$ in (\ref{correl}) can be calculated explicitly,
giving 
\begin{equation}  \label{correl2}
\langle \phi^2 (t) \rangle \, \sim \, {\frac{{m^2 M_P^2} }{{32 \pi^4
\varphi_0^4(t)}}} {\left[ {\frac{{2 \pi \varphi_0^2(t_i)} }{{M_P^2}}} 
\right] }^3 \, .
\end{equation}
Making use of (\ref{tzerolong}), we finally obtain the fractional
contribution of scalar perturbations $\rho_s$ to the total energy density 
\begin{equation}  \label{result}
{\frac{{\rho_s(t)} }{{\rho_0}}} \, \simeq \, - {\frac{{3} }{{4 \pi}}}
{\frac{{m^2 \varphi_0^2(t_i)} }{{M_P^4}}} 
\left[ {\frac{{\varphi_0 (t_i)} }{{\varphi_0 (t)}}} \right]^4 \, ,
\end{equation}
where $\rho_0$ is the background energy density of the homogeneous scalar
field $\varphi_0$.  
In situations in which the ratio (\ref{result}) becomes of the
order $1$, back-reaction becomes very important.

Several consequences can be derived from (\ref{result}). First of all,
back-reaction may lead to a shortening of the period of inflation. Without
back-reaction, inflation would end when $\varphi_0 (t) \, \sim \, M_P$ (see
e.g. \cite{Linde}). Inserting this value into (\ref{result}), one can
expect that if 
\begin{equation}
\varphi_0 (t_i) \, > \, \varphi_{br} \, \sim \, m^{-1/3} M_P^{4/3} \, ,
\end{equation}
then back-reaction will become important before the end of inflation and
may shorten the period of inflation. It is interesting to compare this
value with the scale
\begin{equation}
\varphi_0 (t_i) \, \sim \, \varphi_{sr} \, = \, m^{-1/2} M_P^{3/2} \, ,
\end{equation}
which emerges in the
scenario of stochastic chaotic inflation$^{\cite{Starob,Slava}}$ as 
the ``self-reproduction" scale beyond which quantum fluctuations dominate
the evolution of $\varphi_0 (t)$. Notice that $\varphi_{sr} \gg
\varphi_{br}$
since $m \ll M_P$. Hence, even in the case when self-reproduction does
not take place, back-reaction effects can be very important.

Alternatively, we can fix $\varphi_0 (t_i)$ and use the expression (\ref
{result}) to determine at which value of $\varphi_0$ (denoted by $\varphi_0
(t_f)$) back-reaction becomes important, which presumably implies that
inflation will end at that point. The result is 
\begin{equation}
\varphi_0 (t_f) \, \sim \, {\frac{{m^{1/2} \varphi_0(t_i)^{3/2}} }{{M_P}}}
\, .
\end{equation}

We will conclude this section with an analysis of the back-reaction of
scalar perturbations on the evolution of the homogeneous component of
scalar field $\varphi _0(t)$. The equation for the scalar field $\varphi
_0(t)$
taking into account the backreaction of perturbations can be obtained if
we start with the exact Klein-Gordon equation
\begin{equation}
\dalamb_{g_0+\delta g}(\varphi _0+\delta \varphi
)+V'(\varphi _0+\delta \varphi ) = 0\,\, ,
\end{equation}
expand it to second order in perturbations $\delta g,\delta \varphi $ and
take the average. The result is 
\begin{eqnarray}
&&(\ddot{\varphi_0} \, + 3 H \dot{\varphi_0} ) 
	(1 \, + \, 4 \langle \phi^2 \rangle ) \,
+ \, V' \, + \, \frac{1}{2} V''' \langle \delta\varphi^2 \rangle \,
- \, 2 \langle \phi \delta \ddot{\varphi} \rangle \nonumber \\
&& - 4 \langle \dot\phi \delta \dot\varphi \rangle \, 
- \, 6 H \langle \phi \delta \dot\varphi \rangle \,
+ \, 4 \dot{\varphi_0} \langle \dot\phi \phi \rangle \,
- \, \frac{2}{a^2} \langle \phi \nabla^2\delta\varphi \rangle \, =
\, 0 \,\, .
\label{breq}
\end{eqnarray}
For long
wavelength perturbations in inflationary Universe, the terms containing
spatial or time derivatives of $\phi $ and $\delta \varphi $ are negligible.
Hence, for the potential (\ref{pot}), the  equation (\ref{breq}) becomes 
\begin{equation}
\ddot{\varphi_0} \, + \, 3 \frac{\dot{a}}{a} \dot{\varphi_0}
\, = \, - \frac{m^2}{1 + 4\lphir} \varphi_0 \,\, .
\end{equation}
We conclude that back-reaction of
perturbations  on the evolution of $\varphi _0$ become very important when 
$\langle \phi^2 \rangle \sim 1$, at the same time as they become
important for the evolution of the background geometry of space-time.

\section{Conclusions}

\vskip0.1in

We have defined a gauge invariant  effective energy-momentum tensor (EMT) of
cosmological perturbations which allows us to describe  the back-reaction 
of the perturbations on the evolution of background space-time . Our
formalism can be applied both to scalar and tensor perturbations and applies
independent of the wavelength of the fluctuations.

In particular, our analysis applies to cosmological perturbations produced
during an period of inflation in the very early Universe. In this case, we
have worked out the specific form of the effective EMT for both density
perturbations and gravitational waves. The contribution of long wavelength
scalar fluctuations to the energy density has a negative sign and thus
counteracts the cosmological constant (note that this effect is a purely
classical one, in contrast to the quantum mechanical effects
counteracting the cosmological constant discussed in Refs.
\cite{Ford,Mottola,Woodard}). The equation of state of the
effective EMT is the same as that of the background. The contribution of long
wavelength gravitational waves to the energy density also has a negative
sign, and in this case the equation of state is $p = - {1/3} \rho$. Note that
an instability of de Sitter space to long wavelength fluctuations was
also discovered in Ref. \cite{Deser}.

Applied to the chaotic inflationary Universe scenario, we found that  the
back-reaction of generated perturbations on the evolution of the background
can become very important before the end of the inflation even if we start
at an energy scales below the ``self-reproduction scale''.

\vskip 0.5cm \noindent {\bf Acknowledgements}

Two of us (R.A. and R.B.) wish to thank Fabio Finelli for useful
conversations. This work is supported in part by CNPq (Research Council of
Brazil), Award
20.0727/93.1 (R.A.), by the U.S. DOE under Contract DE-FG0291ER40688, Task A
(R.A. and R.B.), and by the SNF and the Tomalla Foundation (V.M.) The
authors also acknowledge support by NSF Collaborative Research Award
NSF-INT-9312335.


\end{document}